\documentclass[pra,twocolumn,showpacs,amsmath,amssymb,nofootinbib]{revtex4}

\usepackage{graphicx}

\begin{document}
\title{Some remarks on port-based teleportation}
\author{Satoshi Ishizaka}
\affiliation{Graduate School of Integrated Arts and Sciences,
Hiroshima University,
1-7-1 Kagamiyama, Higashi-Hiroshima, 739-8521, Japan}
\date{\today}
%

\begin{abstract}
Port-based teleportation (PBT) is a teleportation scheme such that
the teleported state appears in one of receiver's multiple output ports 
without any correcting operation on the output port.
In this paper, we make some remarks on PBT. Those include the possibility of
recoverable PBT (a hybrid protocol between PBT and the standard
teleportation scheme), the possibility of port-based superdense coding (a dual
protocol to PBT), and the fidelily upper bound expected from the entanglement
monogamy relation in asymmetric universal cloning.
\end{abstract}

%
\pacs{03.67.Hk, 03.67.Ac, 03.67.Bg, 03.67.Lx}
\maketitle
%
\section{Introduction}
\label{sec: Introduction}

Quantum teleportation \cite{Bennett95b} is the most fundamental protocol
in quantum information science, and indeed has always played a crucial role
in the progress of quantum information theory and technology (for a good
review, see \cite{Pirandola15a}).
The standard teleportation scheme (STS) \cite{Bennett95b} transfers an
unknown quantum state from Alice to Bob as follows:
Alice performs a joint measurement on the state to be teleported and half of
the previously shared entangled state, tells the outcome of the measurement
to Bob, and Bob applies a unitary transformation, depending on the outcome,
to the remaining half of the entangled state. To ensure no-signaling
(no faster than light communication), every teleportation scheme must
accompany some sort of communication and Bob's operation depending on the
communication content.

In so-called port-based teleportation (PBT) \cite{Ishizaka08b,Ishizaka09a},
Bob's operation is quite simple: he regards his half of (large) entangled state
as a collection of $N$ output ports, and he only picks up an output port (and
discards all the other ports). The output port contains the teleported
state as it is, without any correcting operation on the output port.
The absence of the correcting operation leads to an
application of PBT as a universal programmable quantum
processor \cite{Nielsen97a,Ishizaka08b}, which is a device to play back the
record of the (past) experiences of a quantum object. The universality of the
device is so powerful that arbitrary experiences can be recoded and played
back (just like a machine in science fiction to relive your childhood),
including not only
those described as unitary evolution but also measurements (working as a
quantum multimeter \cite{Dusek02a} in this case). The drawback is that huge
amount of entanglement is necessary to increase the fidelity or success
probability. It has been shown, however, that the most of the huge amount
of entanglement can be recycled for subsequent PBT \cite{Strelchuk13a}.

Moreover, it has been shown in \cite{Beigi11a} that a combined protocol of
PBT and STS works as if it could break the barrier of spacetime as follows:
Suppose that 
Alice and Bob, separated in spacetime, each has a quantum system $A_0$ and
$B_0$, respectively. Alice can then consider, {\it without waiting for
communication}, that a port of her half of the shared entangled state
contains the state of $A_0B_0$, i.e. she already has the non-local state of
$A_0B_0$ somewhere in her hand (though she can know which port contains the
state only after the communication from Bob) \cite{Beigi11a}. This technique is
used for attacking position-based cryptography and for instantaneous non-local
quantum computation \cite{Beigi11a}. Moreover, PBT has been used as
a tool to investigate the relation of quantum communication complexity and
the Bell non-locality \cite{Buhrman15a}.

However, the properties of PBT have not been completely clarified yet,
in particular, for teleporting a high dimensional quantum state.
This is because, in contrast to STS, the simple multiple use of PBT for a
qubit (quantum bit) does not result in PBT for higher dimension. 
For teleporting a state of a qudit ($d$-dimensional system),
only a lower bound of the teleportation fidelity of deterministic PBT
\cite{Ishizaka09a,Beigi11a} and an upper bound of the success probability of
probabilistic PBT \cite{Garcia13a} have been obtained so far. More studies
will be necessary to clarify the properties of PBT.

In this paper, we make some remarks on PBT. After recalling the formulation
of PBT in Sec.\ \ref{sec: Formulation of PBT}, in 
Sec.\ \ref{sec: Remarks for $d=2$},
we pay attention to the fact that, in most cases of $d=2$, the optimal
measurements of Alice agree with each other.
In Sec. \ref{sec: Recoverable PBT}, we propose a hybrid protocol 
between PBT and STS (say recoverable PBT), where Bob has another choice
(in addition to adopt usual PBT) to adopt a faithful teleportation by
utilizing all the $N$ output ports.
In Sec.\ \ref{sec: Rederivationo of probability bound}, we consider the setting
of the port-based superdense coding, a dual protocol to
PBT, and rederive the upper bound of success probability of probabilistic PBT.
This bound is tight even for $d=3$ and $N=2$.
In Sec.\ \ref{sec: Fidelity bound due to monogamy}, we obtain an upper bound
of the teleportation fidelity by using the entanglement monogamy relation
in asymmetric $1\rightarrow N$ universal cloning.
In Sec.\ \ref{sec: Port-based superdense coding}, we finally remak that
the superdense coding capacity can be asymptotically achieved in a limit
different from the fidelity, and hence port-based superdense coding is
possible. A summary is given in Sec.\ \ref{sec: Summary}.

%
\section{Formulation of PBT}
\label{sec: Formulation of PBT}

To begin with, let us recall the formulation of (deterministic)
PBT, where Bob has $N$ output ports and a teleported state appears
in one of the $N$ ports without any correcting operation on each port.
As a preparation of PBT, Bob has $N$ qudits: $B_1$,
$B_2$, $\cdots$, $B_N$, where each corresponds to the output port of PBT.
In this paper, $B_1$, $\cdots$, $B_N$ are denoted by $B$ as a whole.
Alice also has $N$ qudits: $A_1$, $A_2$, $\cdots$, $A_N$, which are denoted by
$A$ as a whole.
Let us then describe an entangled state between $A$ and $B$ used for PBT as
\begin{equation}
|\Psi\rangle=(O_A\otimes\openone)|\psi^-\rangle_{A_1B_1}|\psi^-\rangle_{A_2B_2} \cdots
|\psi^-\rangle_{A_NB_N}.
\end{equation}
Hereafter, the qudits are regarded as $s$-spins ($d=2s+1$), and $d$ and $2s+1$
will be used interchangeably. The spin basis is denoted by $|s,m\rangle$
($m=-s$, $\cdots$, $s$). Then,
\begin{equation}
|\psi^-\rangle=\frac{1}{\sqrt{2s+1}}\sum_{m=-s}^{s}(-1)^{s-m}|s,m\rangle|s,m\rangle
\end{equation}
is a state of spin 0 in two $s$-spins, which is maximally entangled between the
two. The operator $O$ specifies the actual form of
$|\Psi\rangle$, and $\hbox{tr}O^\dagger O=d^N$ so that $|\Psi\rangle$ is
normalized. Note that, in PBT, the teleportation fidelity is maximized when
$O\ne\openone$ in general, i.e. when $|\Psi\rangle$ is not maximally entangled.

To teleport the state of the $C$ qudit, Alice performs a joint measurement with
$N$ possible outcomes ($1$, $2$, $\cdots$, $N$) on the $A$ and $C$ qudits.
Let us denote the positive operator valued measure (POVM) of her measurement 
by $\{\Pi_i\}$ and hence $\sum_{i=1}^{N}\Pi_i=\openone_{AC}$.
When Alice obtains the outcome $i$, the state of $B_i$ qudit is close to the
state of the $C$ qudit as it is.
It is then found that the entanglement fidelity of PBT is given by
\begin{equation}
F=\frac{1}{d^2}
\sum_{i=1}^{N} \hbox{tr}\Pi_{iAC} [O_A\sigma^{(i)}_{AC}O^\dagger_{A}],
\end{equation}
where
\begin{equation}
\sigma^{(i)}_{AC}=\frac{1}{d^{N-1}}P^-_{A_iC} \otimes \openone_{\bar A_i},
\label{eq: sigma}
\end{equation}
$P^-\equiv|\psi^-\rangle\langle\psi^-|$, and ${\bar A}_i$ denotes the $A$
qudits except for $A_i$ (i.e. $A_1A_2\cdots A_{i-1}A_{i+1}\cdots A_N$).

In analyzing the properties of PBT, the following operator:
\begin{equation}
\rho\equiv\sum_{i=1}^N \sigma^{(i)}=\frac{1}{d^{N-1}}\sum_{i=1}^N 
P^-_{A_iC} \otimes \openone_{\bar A_i}
\label{eq: rho}
\end{equation}
frequently plays a crucial role. Indeed, Alice's optimal measurement in the
case of $d=2$ and $O=\openone$ is the square-root
measurement (SRM) \cite{Ishizaka08b,Ishizaka09a}
[also known as a pretty good measurement (PGM) or least-squares measurement (LSM) \cite{Hausladen94a,Hausladen96a,Ban97a,Sasaki98a,Kato03a,Eldar04a}]
for distinguishing the quantum signals $\{\sigma^{(i)}\}$, and the
corresponding entanglement fidelity is given by
\begin{equation}
F=\frac{1}{d^2}\sum_{i=1}^{N}\rho^{-1/2}\sigma^{(i)}\rho^{-1/2}\sigma^{(i)}.
\end{equation}

To investigate the general properties of $\rho$, let us decompose
$\openone_{\bar A_i}$ into the spin components:
\begin{equation}
\openone_{\bar A_i}=\sum_{j=j_{\rm min}}^{(N-1)s}\openone(j)_{\bar A_i},
\end{equation}
where $\openone(j)$ is an identity on the subspace where the total spin angular
momentum of $\bar A_i$ is $j$ ($j_{\rm min}$ is the possible minimum value).
Then, $\rho$ is also decomposed into
\begin{equation}
\rho(j)\equiv \frac{1}{d^{N-1}}\sum_{i=1}^N 
P^-_{A_iC} \otimes \openone(j)_{\bar A_i}.
\end{equation}
Since the addition of spin 0 and spin $j$ results in spin $j$ only, each term 
$P^-_{A_iC} \otimes \openone(j)_{\bar A_i}$ is an operator on the
subspace of total spin $j$ (though the spin function constructed by the
addition is different, depending on $i$), and hence $\rho(j)$ is also
an operator on the subspace of total spin $j$. Therefore,
$\hbox{tr}\rho(j)\rho(j')=0$ for $j\ne j'$, i.e. $\rho$ is block diagonal with
respect to the total spin angular momentum (and clearly its $z$-component also)
of $N\!+\!1$ spins ($AC$), but the maximum momentum is limited to $(N\!-\!1)s$.

As far as we know, the eigenvalues and eigenstates of $\rho$ for $d>2$ have
not been obtained yet, which leads to the difficulty in analyzing PBT in
higher dimension. The only exception is the maximum eigenvalue, which is proved
in Appendix \ref{sec: Maximum eigenvalue of rho} to be
\begin{equation}
\lambda_{\rm max}=\frac{N+d-1}{d^N}.
\end{equation}
This leads to a known monogamy relation of singlet fraction \cite{Kay09a}
for a multipartite state $\Omega$ such that
\begin{align}
\frac{1}{N}\sum_{i=1}^{N} F_i
&\equiv\frac{1}{N}\sum_{i=1}^{N} \hbox{tr}P^-_{A_iC} \Omega_{AC}  \cr
&=\frac{d^{N-1}}{N} \hbox{tr}\rho\Omega\le
\frac{N+d-1}{Nd},
\end{align}
which explains the fidelity limit of symmetric $1\!\rightarrow\!N$
universal cloning \cite{Werner98a,Scarani05a,Fan14a} in a simpler way.

%
\section{Remarks for $d=2$}
\label{sec: Remarks for $d=2$}

Fortunately, all the eigenvalues and eigenstates of $\rho$ for $d=2$ can be
analytically obtained as we showed in \cite{Ishizaka08b,Ishizaka09a}, but
the results are not simple. Therefore, it may be worth to summarize it again
in a tractable way. A special property held for $d=2$ is that
the spin angular mometum of the $A$ spins (denoted by $S_A$ hereafter)
is a good quantum number, i.e. $\rho$ is block diagonal with respect to
$S_A$ also. Indeed, $\rho(j)$ for $d=2$ is written as
\begin{equation}
\rho(j)=\frac{N-2j+1}{2^{N+1}}\openone(j)^{j-\frac{1}{2}}_{AC}
+\frac{N+2j+3}{2^{N+1}}\openone(j)^{j+\frac{1}{2}}_{AC},
\end{equation}
where $\openone(j)^{j\pm\frac{1}{2}}_{AC}$ is an identity on the subspace where
total spin angular momentum of $AC$ is $j$ and $S_A=j\pm\frac{1}{2}$.
Note that, since the total spin of $j$ is the result of the addition of $S_A$
and $\frac{1}{2}$-spin of $C$,
$\openone(j)=\openone(j)^{j-\frac{1}{2}}+\openone(j)^{j+\frac{1}{2}}$ holds.

When $|\Psi\rangle$ used for PBT is fixed to a maximally entangled state
($O=\openone$), the optimal measurement of Alice to provide the maximum
entanglement fidelity is SRM, i.e. the POVM elements are given by
\begin{equation}
\Pi_i=\rho^{-1/2}\sigma^{(i)}\rho^{-1/2}.
\label{eq: SRM}
\end{equation}
Here, it is implicitly assumed that the excess term $(1/N)\Pi_0$ is added to
every $\Pi_i$ so that $\sum_{i}\Pi_i=\openone$.
Since $j$ in $\rho(j)$ takes a value only for $j\le(N-1)/2$ as mentioned
before, we have
\begin{equation}
\Pi_0=\openone-\openone_\rho=\openone(\textstyle\frac{N+1}{2}),
\label{eq: SRM0}
\end{equation}
where $\openone_\rho$ is an identity on the support of $\rho$.

The corresponding entanglement fidelity can be more increased by optimizing
$O$. The optimization result is
\begin{equation}
O=\sum_{j=j_{\min}}^{N/2}\sqrt{\frac{2^{N+1}}{h^{[N]}(j)(N+2)}}\sin\frac{\pi(2j+1)}{N+2}\openone(j)_A,
\label{eq: Deterministic optimal O}
\end{equation}
where $h^{[N]}(j)$ is the number of states with total spin $j$ in $N$
$\frac{1}{2}$-spins and hence $h^{[N]}(j)=(2j+1)^2\binom{N}{N/2+j}$.
We showed in \cite{Ishizaka09a} the corresponding optimal
measurement of Alice in the form of $\tilde\Pi_i\equiv O^\dagger\Pi_i O$.
However, when the actual POVM elements $\Pi_i$ are derived from $\tilde\Pi_i$,
it will be found that those agree with Eq.\ (\ref{eq: SRM}).
Note that, since $\rho$ is block diagonal with respect to $S_A$ and the
optimal $O$ is an identity on each subspace, we have $[\rho,O]=0$ and as a
result the measurement can also be considered as SRM for distinguishing 
$\{O\sigma^{(i)}O^\dagger\}$.

In the probabilistic version of PBT, the optimal $O$ which provides
the maximum success probability of faithful teleportation $p\!=\!N/(N+3)$ 
is give by
\begin{equation}
O=\sqrt{\frac{2^N}{\sum_j (2j+1)^2}}
\sum_{j=j_{\min}}^{N/2}\sqrt{\frac{(2j+1)^2}{h^{[N]}(j)}}\openone(j)_A.
\label{eq: Probabilistic optimal O}
\end{equation}
We showed in \cite{Ishizaka09a} the corresponding optimal measurement
in the form of $\tilde\Pi_i=P^-_{A_iC}\otimes\tilde\Theta_{i\bar A_i}$, but
it will be found that the actual POVM elements again agree with
Eq.\ (\ref{eq: SRM}) without the implicit excess term of $(1/N)\Pi_0$.
In this probabilistic case, $\Pi_0$ of Eq.\ (\ref{eq: SRM0}) by itself
constitutes a POVM element, such that $\Pi_0$ indicates the failure of
faithful teleportation.

In this way, the optimal measurement of Alice for $d=2$ is given by
Eq.\ (\ref{eq: SRM}) in many cases: both $O=\openone$ and optimal $O$ in the
deterministic version, and optimal $O$ in the probabilistic version. The only
exception is the case of $O=\openone$ in the probabilistic version. This 
seems to rely on the property that $\rho$ is block diagonal with respect to
$S_A$. Unfortunately, this property does not hold for general $d$ as shown in
Appendix \ref{sec: s=1}, where the result of $d=3$ and $N=2$ is explicitly
shown. It is quit interesting that, even in this case, the optimal measurement
of probabilistic PBT again agrees with SRM for distinguishing
$\{\sigma^{(i)}\}$.

%
\section{Recoverable PBT}
\label{sec: Recoverable PBT}

According to the no-go theorem for the faithful and deterministic universal
programmable processor \cite{Nielsen97a}, a deterministic PBT protocol is
inevitably forced to be an approximate one for finite $N$.
Therefore, it may
be convenient if, for the same measurement of Alice, Bob can lately choose
between two choices: (1) usual PBT (with non-unit fidelity) by selecting
one of the $N$ output ports or (2) faithful teleportation (with unit fidelity)
by utilizing all the $N$ output ports. This protocol, say recoverable PBT, is
indeed possible as shown below.

To this end, let us consider the optimal probabilistic PBT for $d=2$ to
teleport the $C$ qubit of $P^-_{CD}$. When Alice obtains the
outcome $\Pi_i$ with $i\ne0$ in her measurement, the state of the $C$
qubit is faithfully teleported to the $B_i$ qubit, and hence the resulting
state is $P^-_{DB_i}$. When Alice obtains $\Pi_0$ that indicates the failure of
faithful teleportation, the state of $BD$ is give by
\begin{align}
\lefteqn{\hbox{tr}_{AC}\Pi_{0AC}O_A(P^-_{CD}\otimes P^-_{A_1B_1}\otimes \cdots \otimes P^-_{A_NB_N})O_{A}} \quad \quad \cr
&=\frac{1}{2^{N+1}}O_B\Pi_{0BD}O_B = \frac{1}{2^{N+1}}O_B \openone(\textstyle\frac{N+1}{2})_{BD}O_B \cr
&=\frac{6\cdot 2^N}{(N+2)(N+3)}
\sum_m |\textstyle\frac{N+1}{2},m\rangle\langle \frac{N+1}{2},m|,
\label{eq: State of BD}
\end{align}
where we used Eq.\ (\ref{eq: SRM0}) and (\ref{eq: Probabilistic optimal O}),
and
\begin{align}
|\textstyle\frac{N+1}{2},m\rangle
&=
\sqrt{\frac{\frac{N}{2}+m+\frac{1}{2}}{N+1}}
|\uparrow\rangle_D|\textstyle\frac{N}{2},m-\frac{1}{2}\rangle_B \cr
&+
\sqrt{\frac{\frac{N}{2}-m+\frac{1}{2}}{N+1}}
|\downarrow\rangle_D|\textstyle\frac{N}{2},m+\frac{1}{2}\rangle_B.
\label{eq: Old basis}
\end{align}
Now, suppose that, after Alice obtains $\Pi_0$, she further measures the
$z$-component of the total spin of $AC$ to determine $m$ in
Eq.\ (\ref{eq: State of BD}). When she obtains $m$, the state of $BD$ becomes
proportional to $|\textstyle\frac{N+1}{2},m\rangle$, which is not maximally
entangled between $D$ and $B$ unless $m=0$ from Eq.\ (\ref{eq: Old basis}).
In this case, the initial entanglement of $C$ is not
transfered to $B$ and Bob cannot recover the lost entanglement anymore.
Instead of this measurement, therefore, suppose that Alice performs the
measurement in the basis of
\begin{equation}
\left\{\begin{array}{ll}
|e^\pm_{m}\rangle\equiv
\frac{|\textstyle\frac{N+1}{2},m\rangle\pm |\textstyle\frac{N+1}{2},-m\rangle}
{\sqrt{2}} & \hbox{~for $m>0$}, \cr
|e_{0}\rangle\equiv
|\textstyle\frac{N+1}{2},0\rangle & \hbox{~for $m=0$}.
\end{array}\right.
\label{eq: New basis}
\end{equation}
It is not difficult to see that, when $N$ is odd and thus $m$ is an integer,
all the above states are maximally entangled between $D$ and $B$, and hence
the entanglement of $C$ is completely transfered to $B$ in this case. This
implies that, if Bob knows the outcome of the measurement
[denoted by $(m,\pm)$] and he applies an
appropriate unitary transformation on $B$ according to the outcome, he can
completely recover the state of $C$ in his hand.
Note that, this does not work well for even $N$, because $|e^\pm_{m}\rangle$
is not maximally entangled for $m=1/2$.

To summarize, the explicit protocol of recoverable PBT is as follows: Alice
performs a measurement $\{\Pi_0,\Pi_1,\cdots,\Pi_i,\cdots,\Pi_N\}$ on $AC$ as in the case
of optimal probabilistic PBT and obtains the outcome $i$. When she obtains
$i\!=\!0$, she further performs the measurement in the basis of
Eq.\ (\ref{eq: New basis}) and obtains $(m,\pm)$. She then send the outcome
$i$ and $(m,\pm)$ to Bob. For $i\ne0$, the state of the $C$ qubit is faithfully
teleported to the $B_i$ qubit. For $i=0$, Bob has two choices. If he ignores
$(m,\pm)$ and randomly picks up one of the $B$ qubits as an output port, the
protocol works as deterministic PBT. The entanglement fidelity is equal to the
probability of obtaining $i\ne0$, because $\hbox{tr}\Pi_0\sigma^{(i)}=0$,
and hence $F=N/(N+3)$. 
If Bob utilizes the information of $(m,\pm)$ to
apply an appropriate unitary transformation to the whole of the $B$ qubits,
he can obtain the state of the $C$ qubit faithfully. The recoverable PBT is
considered to be a
hybrid of PBT and STS. Indeed, the protocol completely agrees with STS for
$N=1$, where $|e^\pm_{1}\rangle=|\phi^\pm\rangle$ and $|e_0\rangle=|\psi^+\rangle$ in the standard notation of the Bell basis.

\begin{figure}[t]
\centerline{\scalebox{0.4}[0.4]{\includegraphics{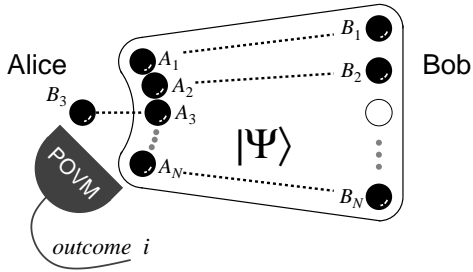}}}
\caption{The setting of port-based superdense coding, where Bob sends the
$B_k$ qudit (the case of $k=3$ is shown in this figure) of previously shared
$|\Psi\rangle$ to Alice, and Alice performs an joint measurement on $B_kA$
to know the value of $k$.
}
\label{fig: PBSDC}
\end{figure}

%
\section{Rederivation of probability bound}
\label{sec: Rederivationo of probability bound}

It has been shown that the success probability of probabilistic PBT for any
$d$ is upper bounded by \cite{Garcia13a}
\begin{equation}
p\le \frac{N}{N+d^2-1}.
\label{eq: Probability bound}
\end{equation}
It seems very plausible that this bound is indeed reachable, because the bound
agrees with the optimal probability for $d=2$ with any $N$, and even for the
case of $d=3$ with $N=2$, where $\rho$ is {\em not} block diagonal with respect
to $S_A$, as shown in Appendix \ref{sec: s=1}.
 Here, we rederive the bound in a way different from
\cite{Garcia13a}, which is convenient for the later discussions.

To this end, let us consider the setting of port-based superdense coding
as shown in Fig.\ \ref{fig: PBSDC}, where Alice and
Bob previously share $|\Psi\rangle$, Bob sends the $B_k$ qudit to Alice,
and Alice performs a measurement on $B_kA$ to know the actual value of
$k$.  Note that the roles of Alice and Bob are opposite to the usual setting
of superdense coding, and note that $|\Psi\rangle$ is not necessarily a
maximally entangled state.
Suppose that, to know $k$, Alice performs the same measurement as
probabilistic PBT, whose POVM elements are $\{\Pi_0,\Pi_1,\cdots,\Pi_N\}$,
and let $q_{i|k}$ be the probability that Alice obtains outcome $i$ ($\ne0$)
when Bob sent $B_k$ to Alice.
Since the state that Alice measures is obtained by projecting
$P^-_{DB_k}$ to
$P^-_{CD}\otimes|\Psi\rangle\langle\Psi|$, we have
\begin{align}
q_{i|k}
&=d^2 \hbox{tr}\Pi_{iAC}(P^-_{CD}\otimes|\Psi\rangle\langle\Psi|)P^-_{DB_k} \cr
&=d^2 \hbox{tr}\left[\hbox{tr}_{AC}\Pi_{iAC}(P^-_{CD}\otimes|\Psi\rangle\langle\Psi|)\right]P^-_{DB_k}.
\end{align}
Here, $\hbox{tr}_{AC}\Pi_{iAC}(P^-_{CD}\otimes|\Psi\rangle\langle\Psi|)$ is
nothing but the post-measurement state in probabilistic PBT to teleport the
half of $P^-_{CD}$, and hence equal to
$p_i (P^-_{DB_i} \otimes \chi_{\bar B_i})$,
where $\chi$ is a normalized state and $p_i$ is the probability that Alice
obtains the outcome $i$ in PBT. We then have
\begin{equation}
q_{i|k}=d^2p_i\hbox{tr}(P^-_{DB_i}\otimes\chi_{\bar B_i})P^-_{DB_k}
=
\left\{\begin{array}{ll}
p_i & \hbox{for $i\ne k$,} \cr
d^2p_k & \hbox{for $i=k$.}
\end{array}\right.
\end{equation}
The success probability of PBT is given by $p=\sum_{i\ne0}p_i$.
Since $\sum_k\sum_{i\ne0} q_{i|k}\le N$ and
\begin{equation}
\sum_k\sum_{i\ne0} q_{i|k}
=\sum_k (\sum_{i\ne0} p_i - p_k+d^2 p_k)
=(N+d^2-1)p,
\end{equation}
we obtain the bound of Eq.\ (\ref{eq: Probability bound}). Note that, in this
derivation,
we only used the fact that the state of $C$ is faithfully teleported to $B_i$
in probabilistic PBT. Note further that $q_{0|k}=0$ must hold so that the
bound of Eq.\ (\ref{eq: Probability bound}) is tight.

%
\section{Fidelity bound due to monogamy}
\label{sec: Fidelity bound due to monogamy}

In the same setting as Fig.\ \ref{fig: PBSDC}, let us now suppose that
Alice performs the same measurement as deterministic PBT. The post-measurement
state, denoted by $\chi_{DB}$ hereafter, is close to but not equal to
$P^-_{DB_i}$. Then, we have
\begin{equation}
q_{i|k}=d^2p_i\hbox{tr}(\chi_{DB})P^-_{DB_k}
=d^2 p_i F_{i|k},
\label{eq: Probablity from fidelity}
\end{equation}
where $F_{i|k}$ is the entanglement fidelity with respect to $P^-_{DB_k}$ when
the state of $C$ is teleported to $B_i$.
For the sake of simplicity, let us consider the symmetric case such that
$p_i=1/N$, $F_{k|k}=F$ (irrespective of $k$), and $F_{i|k}=F'$ for $i\ne k$,
as this permutation symmetry generally holds in PBT.
Namely, $F$ stands for the (usual) entanglement fidelity of the correct output
port, and $F'$ stands for the fidelity of the other output port.
Then, from the condition $\sum_{i\ne0} q_{i|k}=1$ in this deterministic
case, we have
\begin{equation}
F+(N-1)F'=\frac{N}{d^2}.
\label{eq: Singlet equality}
\end{equation}
This equality already implies that faithful and deterministic PBT is impossible
for finite $N$. Indeed, when $F=1$ for the output port $B_i$, the reduced
post-measurement state for the other port $B_k$ must have the form of
$\chi_{DB_k}=(\openone/d)_D\otimes \chi_{B_k}$, and hence $F'=1/d^2$,
but those $F$ and $F'$ cannot satisfy Eq. (\ref{eq: Singlet equality}) for
finite $N$. In this way, Eq. (\ref{eq: Singlet equality}) is a constraint on
entanglement monogamy in PBT. Note that, for $F=1-\epsilon$ with small
$\epsilon>0$,
\begin{equation}
F'=\frac{1}{d^2}-\frac{1}{N-1}(1-\frac{1}{d^2}-\epsilon)<\frac{1}{d^2}
\label{eq: F'}
\end{equation}
and hence $F'$ approaches to $1/d^2$ from below for $N\rightarrow\infty$.

Let us then derive the upper bound of $F$ from the monogamy relation. To this
end, we regard PBT as a kind of asymmetric $1\rightarrow N$ universal
cloning, and consider the monogamy relation derived in \cite{Kay09a,Kay13a,Fan14a}:
\begin{equation}
\sum_{i=1}^N{\cal F}_i\le\frac{d-1}{d}+\frac{1}{N+d-1}\left(\sum_{i=1}^N
\sqrt{{\cal F}_i}\right)^2,
\label{eq: Singlet monogamy}
\end{equation}
where ${\cal F}_i$ is the fully entangled fraction of the $i$-th cloner.
The fully entangled fraction is obtained by maximizing singlet fraction
among local unitary transformations such as
${\cal F}=\max_{UV}\hbox{tr}(U\otimes V)P^-_{AB}(U\otimes V)^\dagger\Omega_{AB}$.
When the following twirling operation is applied to the post-measurement
state $\chi_{DB}$:
\begin{equation}
\int dU (U^*_{D}\otimes U_{B_1}\cdots U_{B_N})
\chi_{DB}(U^*_{D}\otimes U_{B_1}\cdots U_{B_N})^\dagger,
\end{equation}
the resulting reduced states $\chi_{DB_i}$ are all isotropic states,
whose fully entangled fraction has been obtained in \cite{Zhao10a}. We then
have ${\cal F}=F=1-\epsilon$ for the output port $B_i$ and
${\cal F}'=(1-F')/(d^2-1)
=N/[d^2(N-1)]+{\cal O}(\epsilon/N)$ for the other output port $B_k$ because
$F'<1/d^2$ \cite{Zhao10a}. Namely, the fully entangled fraction of
$\chi_{BD_k}$ can take this value, at least.
Putting ${\cal F}$ and ${\cal F}'$ into
Eq.\ (\ref{eq: Singlet monogamy}), we obtain
\begin{equation}
F=1-\epsilon\le 1-\frac{1}{4(d-1)N^2}+{\cal O}(\frac{1}{N^3}).
\end{equation}
In this way, the monogamy relation in asymmetric universal cloning bounds
the entanglement fidelity of PBT
from above by $1-{\cal O}(N^{-2})$. Note that this bound is tight (leaving for
the coefficient) for $d=2$, where
$F=\cos^2\pi/(N+2)\rightarrow 1-\pi^2/(2N^2)$ \cite{Ishizaka09a}.

%
\section{Port-based superdense coding}
\label{sec: Port-based superdense coding}

Superdense coding is a protocol dual to quantum teleportation, where
the classical information capacity of $2\log_2 d$ bits is achieved per qudit
sent from Bob to Alice. In this section, 
we remark that the capacity $2\log_2 d$ bits can be asymptotically achieved,
i.e. port-based superdense coding is possible in the setting of
Fig.\ \ref{fig: PBSDC}.

When Bob sends $B_k$ qudit to Alice, the probability that Alice can obtain
the outcome $i$ by the same measurement as deterministic PBT is
given by Eq.\ (\ref{eq: Probablity from fidelity}).
The entanglement fidelity employing SRM and maximally entangled $|\Psi\rangle$
is lower bounded by $F \ge 1-(d^2-1)/N$ \cite{Ishizaka09a},
but this bound has been slighly improved in \cite{Beigi11a} as
\begin{equation}
F \ge \frac{N}{N+d^2-1}.
\label{eq: Fidelity bound}
\end{equation}
The derivation of this bound using a convenient property of $\rho$,
instead of using $\hbox{tr}\rho^2$, is given in
Appendix \ref{sec: Derivation of fidelity lower bound}.
We then have
\begin{equation}
q_{k|k}=\frac{d^2 F}{N}\ge \frac{d^2}{N+d^2-1}.
\end{equation}
Using this no-error probability, the mutual
information between Bob and Alice, which takes maximum for Bob's equal
prior probability $1/N$, is
\begin{equation}
I(B:A)=\log_2\frac{N}{N+d^2-1}+\frac{d^2}{N+d^2-1}\log_2 d^2.
\end{equation}
At first glance, port-based superdense coding seems impossible because
$I(B\!:\!A)\!\rightarrow\!0$ in the limit of
$d^2\!\ll\!N\!\rightarrow\!\infty$, in quite contrast to 
$F\!\rightarrow\!1$ in the same limit. However, 
$I(B\!:\!A)$ takes the maximum at
$N=(d^2\!-\!1)^2/[(\log_e d^2\!-\!1)d^2+1]$, and therefore with keeping
$N=d^2/\log_e d^2$ in the limit of $N\!\rightarrow\!\infty$, we have
\begin{equation}
I(B:A)\rightarrow 2\log_2 d - \log_2 \log_e d^2.
\end{equation}
In this way, the mutual information asymptotically approaches to the
superdense coding capacity, in the limit different from the fidelity of PBT.
Although the application of
port-based superdense coding is unknown, this may provide an
intriguing example to investigate the duality \cite{Werner01a}
between teleportation and superdense coding.

%
\section{Summary}
\label{sec: Summary}

In this paper, we first recalled the optimal protocols of PBT for $d=2$ and
paid attention to the fact that, in most cases of $d=2$, the optimal
measurements of Alice agree with SRM for distinguishing $\{\sigma^{(i)}\}$.
We showed that, even in the higher dimension of $d=3$, the
optimal measurement of probabilistic PBT for $N=2$ is SRM.
It might be conjectured that this holds for any $d$ and $N$.

Next, we proposed a hybrid protocol between PBT and STS. In this protocol of
recoverable PBT, Bob has two choices, to adopt PBT with an approximate fidelity
by selecting one of $N$ output ports, or to adopt faithful teleportation by
applying a unitary transformation to all the $N$ output ports as STS. We showed
that recoverable PBT is possible at least for $d=2$ and odd $N$.

Moreover, we considered the setting of the port-based superdense coding
as shown in Fig.\ \ref{fig: PBSDC}, and rederived the upper bound of
success probability of probabilistic PBT \cite{Garcia13a}.
In the same setting,
we obtained a constraint between the entanglement fidelities of the output
ports in PBT. We then regarded PBT as asymmetric $1\rightarrow N$ universal
cloning, and derived the upper bound of the fidelity expected from the
entanglement monogamy relation in the asymmetric cloning. The obtained bound
can explain why the entanglement fidelity of PBT is limited to
$F\le 1-{\cal O}(N^{-2})$.

Finally, we remaked that port-based superdense coding is possible. Indeed,
the capacity of $2\log_2 d$ bits per qudit sent is asymptotically achieved
in the limit of $N,d^2\!\rightarrow\!\infty$ with keeping $N=d^2/\log_e d^2$,
while $F\!\rightarrow\!1$ in the limit of $d^2\ll N\rightarrow\infty$ in PBT.
Namely, in spite that port-based superdense coding and PBT are dual to each
other, the perfect transmission of classical and quantum information,
respectively, is achieved in the different limiting conditions. This will be
a good example to deepen our understanding about the duality between
superdense coding and teleportation.

%
\begin{acknowledgments}
This work was supported by JSPS KAKENHI Grants No. 23246071 and No.24540405. 
\end{acknowledgments}
%
\appendix
%
\section{Maximum eigenvalue of $\rho$}
\label{sec: Maximum eigenvalue of rho}

Let $|j,m,\alpha\rangle$ be a spin state of $\bar A_i$ where $\alpha$
distinguishes the permutation degeneracy, and
\begin{equation}
|\xi^{(i)}(j,m,\alpha)\rangle
\equiv|\psi^-\rangle_{A_iC}|j,m,\alpha\rangle_{\bar A_i}.
\end{equation}
By this, the block submatrix of $\rho$ with total spin angular momentum $j$ and
its $z$-component $m$ is written as
\begin{equation}
\rho(j,m)=\frac{1}{d^{N-1}}\sum_{i=1}^{N} \sum_\alpha
|\xi^{(i)}(j,m,\alpha)\rangle\langle\xi^{(i)}(j,m,\alpha)|.
\end{equation}
Let us then define the Gram matrix $\Gamma$ using
$|\xi^{(i)}(j,m,\alpha)\rangle$, i.e. the matrix elements of $\Gamma$ are
given by 
\begin{equation}
\Gamma_{i\alpha,k\beta}=\langle\xi^{(i)}(j,m,\alpha)|\xi^{(k)}(j,m,\beta)\rangle.
\end{equation}
When the spin state $|j,m,\alpha\rangle$ of $\bar A_i$ for $i\!\ne\!1$ is
defined such that
\begin{equation}
|j,m,\alpha\rangle_{\bar A_i}\equiv\big[|j,m,\alpha\rangle_{\bar A_1}\big]_{A_i\rightarrow A_1},
\end{equation}
and hence
$|\xi^{(i)}(j,m,\alpha)\rangle\equiv V_{A_1A_i}|\xi^{(1)}(j,m,\alpha)\rangle$
with $V$ begin a swap operator, it is found that the matrix elements of
$\Gamma$ are given by
\begin{equation}
(2s+1)\Gamma_{i\alpha,k\beta}
=\left\{\begin{array}{l}
(2s+1)\delta_{\alpha\beta}  \hbox{~~~for $i\!=\!k$,}\\
\delta_{\alpha\beta}  \hbox{~~for $i\!\ne\!k$ but $i\!=\!1$ or $k\!=\!1$,}\\
\langle j,m,\alpha|V_{A_iA_k}|j,m,\beta\rangle  \hbox{~~~otherwise.}\\
\end{array}
\right.
\label{eq: Gram matrix}
\end{equation}
Then, $\Gamma$ is a real symmetric matrix because the Clebsch-Gordan (CG)
coefficients are all real. When $\gamma$ is an eigenvalue of $\Gamma$ and
the corresponding normalized eigenvector is 
$\vec c=(\cdots,c_{i\alpha},\cdots)^t$, it is not difficult
to see that $|\psi\rangle=\sum_{i\alpha}c_{i\alpha}|\xi^{(i)}(j,m,\alpha)\rangle$ is an eigenstate of $\rho(j,m)$ and the eigenvalue is $\gamma/d^{N-1}$.
Moreover,
$|\psi\rangle$ is not normalized and $\langle\psi|\psi\rangle=\gamma$.
Now, let us rewrite $|\psi\rangle$ as
\begin{eqnarray}
|\psi\rangle&=&
\sum_i d_i|\psi^-\rangle_{A_iC}\sum_\alpha\frac{c_{i\alpha}}{d_i}
|j,m,\alpha\rangle_{\bar A_i} \cr
&\equiv&\sum_i d_i|\psi^-\rangle_{A_iC}|f^{(i)}\rangle_{\bar A_i},
\end{eqnarray}
where $d_{i}^2=\sum_\alpha c_{i\alpha}^2$, and hence $\sum_id_i^{2}=1$ and
every $|f^{(i)}\rangle$ is normalized. Then,
\begin{eqnarray}
\langle\psi|\psi\rangle&=&
\sum_{ik} d_i d_k 
\langle f^{(i)}_{\bar A_i}|\langle\psi^-_{A_iC}|
\psi^-_{A_kC}\rangle|f^{(k)}_{\bar A_k}\rangle \cr
&=& \sum_i d_{i}^2 + \frac{1}{2s+1}\sum_{i\ne k}d_i d_k
\langle f^{(i)}_{\bar A_i}|\big[|f^{(k)}_{\bar A_k}\rangle\big]_{A_i\rightarrow A_k} \cr
&\le& \sum_i d_{i}^2 + \frac{1}{2s+1}\sum_{i\ne k}|d_i d_k| \cr
&=& \frac{2s}{2s+1}\sum_i d_{i}^2 + \frac{1}{2s+1}(\sum_{i}|d_i|)^2  \cr
&\le& \frac{2s}{2s+1} + \frac{N}{2s+1} = \frac{N+d-1}{d},
\end{eqnarray}
where the Cauchy-Schwarz inequality was used in the second inequality.
As a result, it is found that the maximum eigenvalue of $\rho$ is upper bounded
by $(N+d-1)/d^N$, which is indeed achieved when $d_i=1/\sqrt{N}$ and every
$|f^{(i)}\rangle$ is the same symmetric function, e.g. when $j$ takes the
maximum spin angular momentum $j=(N-1)s$.

%
\section{Optimal probability for $d=3$ and $N=2$}
\label{sec: s=1}

In this case, $\rho$ has only one spin component $j=1$. Let us denote
the spin state on $CA_1A_2$ by $|1,m^{J}\rangle$, which is constructed by
the addition of $S_A=J$ ($J=0,1,2$) and $1$-spin of $C$.
Using the standard relation of the CG coefficients
\cite{Messiah,AngularMomentum}, we have
\begin{align}
\langle 1m^{J}|P^-_{BA_1}|1,m^{J'}\rangle&=\frac{1}{9}(-1)^{J+J'}\sqrt{(2J+1)(2J'+1)}, \cr
\langle 1m^{J}|P^-_{BA_2}|1,m^{J'}\rangle&=\frac{1}{9}\sqrt{(2J+1)(2J'+1)},
\label{eq: P- s=1}
\end{align}
and therefore the matrix elements of $\rho(1,m)$ in the basis of
$|1,m^{J}\rangle$ is
\begin{equation}
\rho(1,m)=
\frac{2}{27}
\left(\begin{array}{ccc}
1 & 0 & \sqrt{5} \cr
0 & 3 & 0 \cr
\sqrt{5} & 0 & 5
\end{array}\right),
\label{eq: rho s=1}
\end{equation}
where the rows and columns are indexed by $J$.
Now, by choosing $O$ as
\begin{equation}
X=OO^\dagger=\frac{6}{5}\openone(0)_A+\frac{3}{5}\openone(1)_A+\frac{6}{5}\openone(2)_A,
\end{equation}
the matrix elements of $I_C\otimes X_A$
in the basis of $|1,m^{J}\rangle$ are
\begin{equation}
\openone_C\otimes X_A=
\frac{1}{5}
\left(\begin{array}{ccc}
6 & 0 & 0 \cr
0 & 3 & 0 \cr
0 & 0 & 6
\end{array}\right),
\label{eq: X s=1}
\end{equation}
because $\openone_C\otimes\openone(J)_A=\sum_{k=|J-s|}^{J+s}\openone^{J}(k)_{AC}$. Then, the POVM elements of 
\begin{equation}
\tilde\Pi_i=O\Pi_i O^\dagger=\frac{9}{10}P^-_{CA_i}\otimes \openone_{\bar A_i}
\end{equation}
satisfy the constraints of probabilistic PBT \cite{Ishizaka09a} cvas
$\hbox{tr}X=9$ and
$\sum_i\tilde\Pi_i=(27/10)\rho\le\openone_C\otimes X_A$.
The corresponding success probability is $p=(1/3^3)\sum_i\hbox{tr}\tilde\Pi_i
=1/5$, which agrees with the upper bound of Eq.\ (\ref{eq: Probability bound}).
It is intriguing that, even in this case, we have
from Eq.\ (\ref{eq: P- s=1}), (\ref{eq: rho s=1}) and (\ref{eq: X s=1}),
\begin{align}
\Pi_i
&=(\openone_C\otimes X_A)^{-1/2}(\frac{10}{9}P^-_{BA_i})(\openone_C\otimes X_A)^{-1/2} \cr
&=\rho^{-1/2}(\frac{1}{3}P^-_{BA_i})\rho^{-1/2}=\rho^{-1/2}\sigma^{(i)}\rho^{-1/2},
\end{align}
and hence the optimal measurement is SRM. Note that $[\rho,O]=0$ also holds,
and
\begin{equation}
\Pi_0=\openone-\openone_\rho=\openone(N+1)+\sum_m |\kappa_m\rangle\langle\kappa_m|,
\end{equation}
where $|\kappa_m\rangle=(5|1,m^{J=0}\rangle-|1,m^{J=2}\rangle)/\sqrt{6}$ is
the eigenstate with a zero eigenvalue of $\rho(1,m)$

%
\section{Derivation of fidelity lower bound}
\label{sec: Derivation of fidelity lower bound}

For the operator $\rho$, the following convenient relations hold:
\begin{align}
\langle\psi^-_{A_iC}|\rho|\psi^-_{A_iC}\rangle&=\frac{N+d^2-1}{d^{N+1}}\openone_{\bar A_i}, \cr
\langle\psi^-_{A_iC}|\openone_\rho|\psi^-_{A_iC}\rangle&=\openone_{\bar A_i}.
\end{align}
By using $X^{-1/2}\ge (3/2) \openone_{X}-(1/2)X$ for $X\ge0$,
and by using $|\xi^{(i)}(j,m,\alpha)\rangle
\equiv|\psi^-\rangle_{A_iC}|j,m,\alpha\rangle_{\bar A_i}$ defined in
Appendix \ref{sec: Maximum eigenvalue of rho}, we have
\begin{align}
F&=\frac{1}{d^2}\hbox{tr}\sum_{i=1}^N \rho^{-1/2} \sigma^{(i)} \rho^{-1/2}
\sigma^{(i)} \cr
&=\frac{Na}{d^{2N}} \sum_{jm\alpha}\left|
\langle \xi^{(1)}(j,m,\alpha)|(a\rho)^{-1/2}|\xi^{(1)}(j,m,\alpha)\rangle
\right|^2 \cr
&\ge\frac{Na}{d^{2N}} \sum_{jm\alpha}\left|
\langle \xi^{(1)}(j,m,\alpha)|[\frac{3}{2}\openone_\rho-\frac{a}{2}\rho]|\xi^{(1)}(j,m,\alpha)\rangle \right|^2 \cr
&=\frac{Na}{d^{2N}} \sum_{jm\alpha}\left|
\langle j,m,\alpha|[\frac{3}{2}\openone-\frac{a(N+d^2-1)}{2d^{N+1}}\openone]|j,m,\alpha)\rangle \right|^2 \cr
&=\frac{Na}{d^{N+1}}\left(\frac{3}{2}-\frac{a(N+d^2-1)}{2d^{N+1}}\right)^2,
\end{align}
where see Eq.\ (\ref{eq: Gram matrix}) for the second equality.
This lower bound is maximized when $a=d^{N+1}/(N+d^2-1)$, and hence we
obtain Eq.\ (\ref{eq: Fidelity bound}).


%

\end{document}